# Purging of Untrustworthy Recommendations from a Grid

P. Suresh Kumar[1], P. Sateesh Kumar[2] and S. Ramachandram[3]

[1]Department of MCA, KITS, Warangal
peddojusuresh@gmail.com
[2]Department of E&C E, IIT, Roorkee
sateesh@ieee.org
[3]Department of CSE, OU, Hyderabad

## ABSTRACT

*In grid computing, trust has massive significance. There is lot of research to propose various models in providing trusted resource sharing mechanisms. The trust is a belief or perception that various researchers have tried to correlate with some computational model. Trust on any entity can be direct or indirect. Direct trust is the impact of either first impression over the entity or acquired during some direct interaction. Indirect trust is the trust may be due to either reputation gained or recommendations received from various recommenders of a particular domain in a grid or any other domain outside that grid or outside that grid itself. Unfortunately, malicious indirect trust leads to the misuse of valuable resources of the grid. This paper proposes the mechanism of identifying and purging the untrustworthy recommendations in the grid environment. Through the obtained results, we show the way of purging of untrustworthy entities.*

## 1. INTRODUCTION

Grid computing has emerged as an important and popular new field as an extension of distributed computing with more focus on high performance computing and resource sharing [1]. Modern science and technology, such as High Energy and Nuclear Physics, Astronomy, Climate and Materials science, are increasingly collaborative and span wide disciplinary and geographical areas. They often demand huge resources of computing, storage, and instruments, which individual research institutes could not possess [2]. With the success of World Wide Web, more emphasis in research was on providing scientific collaborations and cross organizational collaborations including various application providers, service providers, storage providers, business environments with B2B partners. The widely deployed Internet links geographically distributed resources and makes resource and data sharing possible. Security is one of the key issues in such resource and data sharing environment in a grid. In particular, all the nodes contained in the grid may not be trustworthy. Trust has been defined as *an assured reliance on the character, ability, or strength on someone or something* [3]. Some of the nodes may be fraudulent or malicious. Resource sharing or having transactions in such an unpredicted environment may lead to adversity. Reliability of the grid transactions is decided by the trustworthiness of the nodes that interact in the grid. Majority of the recent research in grid computing is focused on identification of trustworthy nodes in the grid. This paper aims to identify the untrustworthy transactions in the grid environment and proposes the method to eliminate them in order not to participate in the transactions.

Many approaches have been proposed to deal with trust management [3]. Traditional approaches of establishing a trust relationship are regulation, self-regulation, third-party certification, and security technology. Inadequacies in traditional techniques lead to the evolvement of reputation-based approach for trust management. Reputation, being a social concept, has been defined as *the confidence in the ability of a specific subject to fulfil a certain task*. An extended study has been done by many researchers to apply the concept of reputation to establish trust among the communicating parties [4]. This paper will use the reputation-based approach of trust management in its proposed method. Trustworthiness in the proposed model is evaluated based on direct experience (direct trust) encountered on the service provider by the service initiator and the reputation received due to feedback or recommendations from the fellow service providers or service initiators. A method is proposed to eliminate the unreliable feedbacks from the reputation information received by the service initiator.

Remaining parts of the paper is organized as follows: Section 2 presents recently reported related work in the literature. Proposed technique of eliminating the untrustworthy grid transactions is described in Section 3. In Section 4, various tuning factors chosen for simulation of the grid are provided and results are produced and analyzed. Finally, Section 5 concludes the paper.

## 2. RELATED WORK

Various models have been proposed using reputation trust in grid environment [5]. Some of the recent models are discussed in this section.

Vivekananth [6] proposed a behavior based trust model which shows the behavior conformity and concentrated on behavior of entities in different domains, in different contexts. The total trust is calculated using direct trust and indirect trust. The behaviour was tracked using a tracking module. Based on experiences with the entities, an entity trust level is increased or decreased. A penalty factor is levied for malicious behavior. The trust factor between two entities may depend on penalty, context and time. The penalty factor ranges between 0 and 1. A threshold value is used and if the total trust is greater than the required trust then the resource is allocated. In our proposed method, we have adapted the similar idea of penalizing the untrustworthy entities.

Srivaramangai et al [7] proposed a trust model to improve reliability in grid. According to their model, reputation based systems can be used in grid to improve the reliability of transactions and reliability is achieved by establishing mutual trust between the initiator and the provider. Indirect trust is taken as the measure from the reputation score of other entities. Unreliable feedbacks are eliminated using Spearman's rank correlation method. In our proposed model, we have adapted the usage of rank correlation method to eliminating the untrustworthy entities. However, we have adapted Kendall's rank correlation method instead of Spearman, as it is proved to be the better method in such typical environments.

Wang Meng et al [8] proposed a Dynamic Grid Trust Model named DyGridTrust which is based on recommendation credibility. This model suggested a way to distinguish honest and dishonest recommendation and adjust the weight of trust evaluation dynamically. This model defines various participating nodes in the grid as sponsor node, goal node and recommended node. In our proposed model, we have adapted the way of giving the weights and credibility to the entities.

Gao Ying et al [9] proposed a layered trust model based on behavior to enhance grid security and extensibility. This model is based on the problem in open service grids to establish trust relationship among different domains. The authors have proposed an algorithm to adjust trust relationships between domains based on entities interactions and also proposed a technique to process recommendation trust. In our proposed model, we have adapted the usage of domains.

Various other models are also proposed which are somehow connected to the models discussed earlier. Some of them are briefed. Kai Wei Shaohua Tang [10] proposed a multi-level trust evaluation model based on direct search in which grid service providers need to evaluate and manage the trust of all users. Tie-Yan Li et al [11] proposed a two-level trust model in which the upper level defines the trust relationships among virtual organizations in a distributed manner and the lower level justifies the trust values within a grid domain. Yuan Lin et al [12] designed a model by adding asymmetric users' behaviors reflecting users' characteristics (both the subjective and objective). Huang Wenming et al [13] studied the characteristics of the two classes of true and false recommendation. Shashi Bhanwar et al [14] proposed a trust model for by computing reputation and trustworthiness of the transacting domain on the basis of number of past transactions and rated feedback score.

## 3. PROPOSED MODEL

The proposed model aims to evaluate the trust relationship between two entities based on either due to information directly available about the other entity and/or due to the reputation information received from different types of entities present in that domain or grid.

### 3.1. Architecture

We have considered three cases in the proposed model.

i. *Intra-Domain Intra-Grid Environment:* An entity is interested to utilize the resources present in that domain in which it is residing. In this domain, all the entities will follow the same access controls. This environment is presented in Figure l.

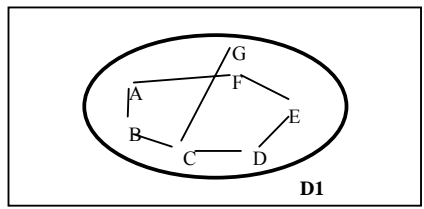

*Figure 1: Intra-Domain Intra-Grid Environment*

ii. *Inter-Domain Intra-Grid Environment:* An entity is interested to utilize the resources present in different domain than in which it is residing, however, in the same grid. In this situation, the domains will have different access controls. This environment is presented in Figure 2.

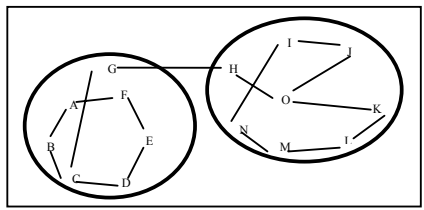

*Figure 2: Inter-Domain Intra-Grid Environment*

iii. *Inter-Grid Environment:* An entity is interested to utilize the resources present in different domain of different grid than in which domain of a grid it is residing. In this situation, the domains may have similar or different access controls but are part of different grids. This environment is presented in Figure 3.

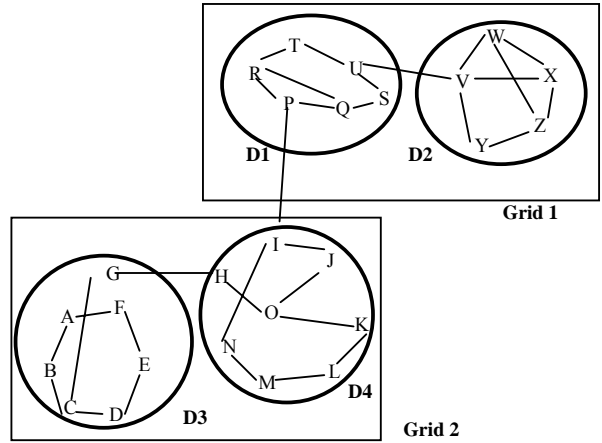

*Figure 3: Inter-Grid Environment*

### 3.2. Evaluation of Trust

The total trust level of an entity is the sum of direct trust and indirect (recommendation/reputation) trust received due to the previous interactions. Evaluation of trust in this model has been explained for the architecture discussed in Section 3.1, as per the following procedure.

### 3.2.1. Direct Trust

For a specific context $C_i$ (one of the three situations of a grid), domain $D_i$ can utilize resources or deploy services using domain $D_j$'s resources and hence a direct relationship will exist between these two domains. Since a direct trust relationship is asymmetric, each of these two domains involved in this direct relationship will have its own interpretation of how well or how bad this direct trust relationship is. Direct trust has been evaluated based on first impression or direct experience with the opponent party and usually it is rated between 0 and 1. A value 0 represents a very poor trust and a value 1 represents an extremely high trust. There are many variations in adapting initial reputation such as based on majority behavior, and assigning the initial reputation based on default value or evaluation of new comer by other grid providers. However, selecting the default value for a new comer is proved to provide best results [7] in terms of fairness and accuracy. A matter of question is what default value should be assigned. Assigning the value 0 may never give a chance for the new host to be allocated as the existing hosts with high first impression will be chosen always. Assigning a value 1 may be a case is likely that a malicious node is given a chance. A good choice of selecting the default value would be an average. For further interactions direct trust can be taken as its previous reputation value which was calculated in last interaction.

### 3.2.2. Reputation Trust

When an entity in domain Di wants to have an interaction with another entity in domain Dj, in addition to the direct trust relationship, Di can rely on recommendations from other domains about Dj (i.e., asking for the reputation of Dj) either from same or different domain of same grid or from different grid. Reputation of an entity in Dj can be estimated based on the recommendations of the entities from the other domains of the same or different domains in the same grid or different. The reputation trust value is calculated based on the following formula i.e., equation (1):

$$R_y(x) = w_1 * DT + w_2 * ITSD + w_3 * ITOD \qquad \ldots\ldots (1)$$

Where  DT    - Direct Trust
       ITSD  - Indirect Trust for Same Domain
       ITOD  - Indirect Trust for Other Domain

Where $w_1$, $w_2$, $w_3$ are weighting factors for the respective reputation of Y with respect to direct trust of X, reputation of Y with respect to recommenders of X in the same domain, and reputation of Y with respect to recommenders of X in the other domain. These factors are empirically determined constants which should satisfy the constraint $w_1 > w_2 > w_3$. Note that reputation values are restricted to values between 0 and $\mu$, where $\mu$ is the pre-defined maximum reputation value, such that $0 \leq R_y(X) \geq \mu$. To achieve this condition the constraint $w_1 + w_2 + w_3 = 1$ should be satisfied.

$$\text{ITSD} = \frac{\sum_{i \neq k} \Phi_i\, R_y(x_i)}{\sum_{i \neq k} \Phi_i} \quad \ldots\ldots (2)$$

$$\text{ITOD} = \frac{\sum_{j \neq k} \Omega_j\, R_y(x_j)}{\sum_{j \neq k} \Omega_j} \quad \ldots\ldots (3)$$

Where $\Phi$, $\Omega$ are credibility factors

In equation (2), $\sum \Phi_i\, R_y(x_i)$ represents the weighted sum of reputations of Y as reported by the X's recommenders ($Xi$) in the same domain.

In equation (3), $\sum \Omega_j\, R_y(x_j)$ represents the weighted sum of reputations of Y as reported by the X's recommenders ($Xi$) in the other domain.

*Calculation of credibility factors:*

In our model, every reputation value, which is obtained from another host, is multiplied by the corresponding credibility factors. These credibility factors represent the trust in the capability of a host to give a valid and dependable recommendation about other hosts. The credibility factors depend on the similarity, the activity, popularity of a certain host which are shown in the equations (5), (6) and (7). The value of $\Phi$ (Same for $\Omega$) is calculated as follows:

$$\Phi = v_1 * \text{similarity} + v_2 * \text{activity} + v_3 * \text{popularity} \quad \ldots\ldots (4)$$

Where $v_1$, $v_2$ and $v_3$ are factors to give the relative importance of a specific parameter with respect to others in equation (4). These values are host-specific and have to be consistently used in all the calculations of the weighting factors.

*Similarity:*

The similarity value determines the similarity of two hosts in their evaluation procedures and their reputation values. The more similar two hosts are, the more credible their recommendation will be with respect to each others. **Kendall's rank correlation method** is used to find the similarity between recommendations. The similarity value between two host recommendations is calculated as follows in equation (5):

$$\text{Similarity} = 1 - \frac{2 * [d_\Delta (s1, s2)]}{n(n-1)} \quad \ldots\ldots (5)$$

In order to compare two ordered sets (on the same set of objects), the approach of this similarity is to count the number of different pairs between these two ordered sets. This number gives a distance between sets called the symmetric difference distance (the symmetric difference is a set operation which associates to two sets the set of elements that belong to only one set). The symmetric difference distance between two sets of ordered pairs s1 and s2 is denoted by $d_\Delta(s1, s2)$. And *n* is the number of hosts present in the domain.

Based on the similarity between two host recommendations, if one host is giving different recommendations compared to other hosts, then it is giving untrustworthy information. So, in next visit this untrustworthy entities can be purged and reliable transactions can be done with the help of this model.

*Activity:*

The activity value reflects the level of activity of a certain host in the past interval of time $t_1$. The more active a host is, the more up-to-date and accurate are its reputation values from its direct experiences and from its received reputation values. Activity is calculated as the fraction of interactions a host performed in the past $t_1$ with respect to the other hosts, of all interactions; however, to keep hosts from "lying", we will calculate the total number of interactions of a host from the other hosts' reputation vectors.

$$\text{Activity} = \frac{\text{Number of interactions by recommenders of a host X}}{\text{Total number interactions by all recommenders}} \quad \ldots\ldots (6)$$

*Popularity:*

The popularity value is a measure of how much a host is liked and how much its services are asked for in the system. The popularity of a host is calculated as the fraction of interactions other hosts have done with this specific host, of all interactions:

$$\text{Popularity} = \frac{\text{Number of interactions with initiator}}{\text{Total number interactions with all other hosts}} \quad \ldots\ldots (7)$$

Based on the above formulas mentioned in equations (1), (2), (3) and (4), the reputation of the host can be calculated. If the calculated reputation is greater than the minimum threshold value the job will be assigned to *y*. Otherwise it will be rejected. After the transaction is over the reputation table will be updated by taking the new value. The decaying factor, mentioned in equation (8) is considered for modifying the reputation of each entity with time.

*Effect of decaying factor:*

As time passes by, a host reputation with respect to other hosts typically changes to an unknown state if little or no interaction occurs between them. When a Host *Z* receives a request (from Host *X*) for reputation information about Host *Y*, it modifies its reputation information relative to *Y* by using decaying factor and then sends the result to the requesting host.

$$R_y(z) = \text{final value} + (\text{final value} - \text{initial value}) * \tau \quad \ldots\ldots (8)$$

Where $\tau$ decay factor which varies is based on time factor. If *t* is the current time and $t_0$ is the time at which the last transaction taken place then the calculation of $\tau$ is as follows.

$$\begin{aligned}
\tau &= 1 & &\text{if } t - t_0 < 1 \text{ month} \\
\tau &= 0.75 & &\text{if } 1 < t - t_0 < 2 \\
\tau &= 0.5 & &\text{if } 2 < t - t_0 < 3 \\
\tau &= 0 & &\text{if } t - t_0 > 3
\end{aligned}$$

*Effect of Threshold values:*

The step that follows the calculation of the reputation of a certain host is to determine trust, by associating to the host the label "trustworthy" or "untrustworthy". This can be determined by introducing two threshold values η and ξ, referred to as the absolute trust and absolute mistrust thresholds, respectively. Three cases are considered:

- If $R_Y(X) \geq \eta$  Y can be trusted
- If $R_Y(X) \leq \xi$  Y cannot be trusted
- If $\xi \geq R_Y(X) \geq \eta$  Y can be considered as either trustworthy or untrustworthy depending on how paranoid or trusting host X is.

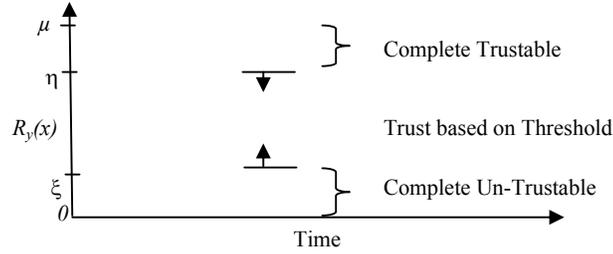

*Figure 4: Threshold values based on reputation values*

## 4. RESULTS

The experimentation is done to purge the untrustworthy entities from the grid. We have considered two grids, 4 domains and 15 entities. Among 15 entities some entities are giving unreliable information which causes wrong allocation of resources to an entity. We have taken 15 entities and same experiment is executed 10 times, which gives variation in the results. In this experiment we have found some entities are behaving maliciously and giving untrustworthy information. We have shown our results in Table 1, in which entities E, J are giving untrustworthy information. Entities E and J are malicious and because of their malicious nature, these entities are giving different results compared to other entities.

We have compared our results with the previous model [7], in which new model identified the untrustworthiness of the entities E and J, where existing model accepting E and J. These two entities E and J can be purged to get reliable transactions and to allocate resources for trustworthy entities only. So, based on the results obtained, our model is giving better results compared to existing model.

*Table 1: comparison of new model with existing model*

| Initiator | Provider | TS1 | Existing Model | TS2 | Proposed Model |
|---|---|---|---|---|---|
| B | I | 1.827 | NO | 1.505 | NO |
| C | E | 2.395 | YES | 1.981 | NO |
| C | I | 1.885 | NO | 0.79 | NO |
| D | N | 1.852 | NO | 1.407 | NO |
| H | G | 2.248 | YES | 2.696 | YES |
| H | N | 2.337 | YES | 2.725 | YES |
| I | M | 1.588 | NO | 1.438 | NO |
| J | M | 2.751 | YES | 1.147 | NO |
| M | F | 1.761 | NO | 1.643 | NO |
| N | A | 2.455 | NO | 2.476 | NO |

## 5. CONCLUSION

In this paper, we proposed a model for trust evaluation in grid environment which will purge the untrustworthy transactions. This model is a reputation-based trust evaluation model having direct and indirect (recommendation/reputation) trust values. In our model, unlike other models, we have considered feedbacks from various *domains* like intra-domain, inter-domain in intra-grid and also from inter-grid. We have also adapted a better ranking algorithm, *Kendall's rank order correlation algorithm*, is used to select the nodes with higher reputation. This way always the nodes with bad feedback are quarantined for a particular period of time. Results indicate that the trust evaluation becomes more robust in this model. This model is giving better results compared to the existing model. The future work is aimed at comparing the same model with other existing models. It is also aimed to evaluate the model in various other aspects to check its robustness.